\documentclass[preprints,article,accept,moreauthors,pdftex,10pt,a4paper]{Definitions/mdpi}
\usepackage{natbib}            % added to moke arXiv submission work
\firstpage{00}
\pubvolume{00}
\issuenum{00}
\articlenumber{0}
\pubyear{2019}
\copyrightyear{2019}
\history{Received: 13 May 2019; Accepted: 8 July 2019; Published: date}
\updates{yes} % If there is an update available, un-comment this line

\newcommand{\lesssim}{\stackrel{\textstyle <}{_\sim}}
\usepackage{booktabs}
\usepackage{multirow}
\usepackage{soul}
\usepackage{microtype}
\usepackage[utf8]{inputenc}
\usepackage{environ}
\NewEnviron{myequation1}{%
\begin{equation}
\scalebox{0.95}{$\BODY$}
\end{equation}}

\NewEnviron{myequation2}{%
\begin{equation}
\scalebox{0.95}{$\BODY$}
\end{equation}}

\NewEnviron{myequation3}{%
\begin{equation}
\scalebox{0.95}{$\BODY$}
\end{equation}}
\pdfoutput=1

\Title{Phases of Hadron-Quark Matter in (Proto) Neutron~Stars}

\Author{  {Fridolin Weber} $^{1,2,*}$, Delaney Farrell $^{1}$,
   {William M. Spinella}  {Milva~G.~Orsaria} $^{4,5}$,  {Gustavo
    A. Contrera} $^{4,6}$ and Ian Maloney $^{1}$}

\AuthorNames{Delaney Farrell, William M. Spinella, Fridolin Weber,
  Germán Malfatti, Gustavo A. Contrera, and Milva G. Orsaria}

\address{$^{1}$ \quad Department of Physics, San Diego State
  University, San Diego, CA 92182, USA\\
  $^{2}$ \quad  {Center for Astrophysics and Space Sciences},
  University of California at San Diego , La Jolla, CA 92093, USA\\
$^{3}$ \quad Department of Sciences, Wentworth Institute of
  Technology, 550 Huntington Avenue, Boston, MA~02115,~ {USA}
 \\
$^{4}$ \quad Grupo de Gravitaci\'on, Astrof\'isica y Cosmolog\'ia,
  Facultad de Ciencias Astron{\'o}micas y Geof{\'i}sicas, Universidad
  Nacional de La Plata, Paseo del Bosque S/N,  La Plata 1900,
  Argentina\\
$^{5}$ \quad CONICET, Godoy Cruz 2290,  {Buenos Aires 1425},
  Argentina\\
$^{6}$ \quad IFLP, UNLP, CONICET, Facultad de Ciencias Exactas,
  Diagonal 113 entre 63 y 64,  La Plata 1900, Argentina\\ }
\corres{Correspondence: fweber@sdsu.edu or fweber@ucsd.edu }

\abstract{In the first part of this paper, we investigate the possible
  existence of a structured hadron-quark mixed phase in the cores of
  neutron stars.  This phase, referred to as the hadron-quark pasta
  phase, consists of spherical blob, rod, and slab rare phase
  geometries.  Particular emphasis is given to modeling the size of
  this phase in rotating neutron stars.  We use the relativistic
  mean-field theory to model hadronic matter and the non-local
  three-flavor Nambu--Jona-Lasinio model to describe quark
  matter. Based on these models, the hadron-quark pasta phase exists
  only in very massive neutron stars, whose rotational frequencies are
  less than around 300 Hz. All other stars are not dense enough to
  trigger quark deconfinement in their cores.  Part two of the paper
  deals with the quark-hadron composition of hot (proto) neutron star
  matter.  To this end we use a local three-flavor
  Polyakov--Nambu--Jona-Lasinio model which includes the 't Hooft
  (quark flavor mixing) term. It is found that this term leads to
  non-negligible changes in the particle composition of (proto)
  neutron stars made of hadron-quark matter.}


\begin{document}

\section{Introduction}\label{sec:introduction}

Neutron stars are extremely compact objects.  They are typically about
20 to 30 km across and spin rapidly, often making many hundreds of 
rotations per second (pulsar PSR J1748-2446ad is the fastest spinning
neutron star known to date, rotating at 716 Hz.)  Depending on mass
and rotational frequency, gravity compresses the matter in the core
regions of neutron stars to densities that are several times higher
than the density of atomic nuclei.  At~such enormous densities, atomic~nuclei are squeezed so tightly together that new types of particles
and novel states of matter may be created.  These include the creation
of hyperons and delta isobars and the formation of a mixed phase of
hadrons and quarks deep in the central regions of neutron stars. As~shown by Glendenning~\cite{Glendenning:1992pt,Glendenning:2001pr}, the~quark-hadro mixed phase region could segregate phases by net charge
to minimize the total energy of the system, leading to the formation
of a hadron-quark Coulomb lattice. Since this lattice is qualitatively
similar to the hypothesized ``nuclear pasta'' structures in the
crustal regions of neutron stars
~\cite{Glendenning:1992pt,Glendenning:2001pr,Lamb:1981kt,Ravenhall:1983,Williams:1985},
it is referred to as hadron-quark pasta phase.\footnote{This phase
  has nothing to do with the crystalline quark matter phases that have
  been introduced in the context of discussing color superconducting
  quark matter phases
~\cite{Rajagopal:2001Handbook,Alford:2001Crystalline}.}  All these
possibilities make neutron stars and the events creating them (e.g.,
GW 170817) superb astrophysical laboratories for a wide range of
physical studies. With~observational data accumulating rapidly
from both orbiting (HST, ATHENA, NICER) and ground based (skA, FAST,
LIGO, VIRGO) observatories there has never been a more exciting time
than today to study neutron~stars.

\section{Hadronic Composition of Neutron Star Matter at Finite~Temperature}
\label{sec:nucl_eos}

A neutron star is born as a result of the gravitational collapse of
the iron core of a massive evolved progenitor star ($M\sim
8$--$25\,M_{\odot}$) in a type-II supernova explosion. The~iron core
of such a star collapses when its mass reaches the Chandrasekhar
limit, which depends on the mass of the progenitor star. Due to the
Fermi pressure of the nucleons, the~collapse stops when nuclear matter
density is reached and the core bounces back.  Shortly after core
bounce (some 10 ms) a hot, lepton rich neutron star, called a
proto-neutron star, is formed
~\cite{Prakash:1997a,Pons:1999ApJ,Steiner:2000a,Prakash:2001Springer,Burgio:2008:a,Malfatti:2019}. This
star consists of a shocked envelope with an entropy per baryon of $s
\sim$ 4--10 (in units of the Boltzmann constant) and an unshocked core
with $s \sim 1$.  The~envelope and the core contain nearly the same
mass of about 0.6--0.8\,$M_{\odot}$.  During~the so-called
Kelvin-Helmholtz cooling phase the lepton number decreases in the
proton-neutron star due to the loss of neutrinos and the proto-neutron
star evolves in about 10 to 30 s to a hot, lepton poor neutron star
with an entropy per baryon of $s \sim$ 1--2.  After~several minutes,
this hot neutron star cools to a cold neutron star with temperatures
$T < 1$ MeV, which then continues to slowly  {cool} via neutrino and
photon emission until its thermal radiation becomes too weak to be
detectable with X-ray telescopes, after~about $10^7$~years.

To study hot hadronic matter at conditions that exist in the cores of
(proto) neutron stars, we use the relativistic nuclear mean-field
(RMF) approximation and choose the  {GM1L, HV, and~DD2} nuclear
parametrizations  {listed in Table}~\ref{table:parametrizations}
that satisfy current constraints from studies of symmetric nuclear
matter and from observations of neutron stars.  The~GM1L
parametrization is based on the standard RMF model and was proposed in
Refs.\ \cite{Glendenning:1992pt,Spinella:2017thesis} but has been
constructed to satisfy recent constraints from symmetric nuclear
matter at saturation density while simultaneously being able to
satisfy the $2.01\, M_{\odot}$ mass constraint set by PSR J0348+0432
with the inclusion of hyperons in SU(3) flavor symmetry.  The~DD2
parametrization is also based on the standard RMF approach but
utilizes a density dependent meson-baryon coupling scheme, with~the
parameters of the density dependence set such that the model
reproduces the properties of finite nuclei
~\cite{Hofmann:2001a,Lenske:1995a,Fuchs:1995DD,Typel:1999DD}. The~lagrangian
of the density-dependent relativistic mean-field model is given
by~\cite{Fuchs:1995DD,Typel:1999DD,Typel:2018DD,Mellinger:2017U}\vspace{-6pt}
\begin{myequation1}
\begin{array}{ccl}
  \mathcal{L} &=& \sum_{B}\bar{\psi}_B \bigl[\gamma_\mu [i\partial^\mu
      - g_{\omega B}(n) \omega^\mu - g_{\rho B}(n) {\boldsymbol{\tau}}
      \cdot {\boldsymbol{\rho}}^\mu] - [m_B - g_{\sigma B}(n)\sigma]
    \bigr] \psi_B + \frac{1}{2} (\partial_\mu \sigma\partial^\mu
  \sigma - m_\sigma^2 \sigma^2) \\
  & -& \frac{1}{3} \tilde{b}_\sigma
  m_N [g_{\sigma N}(n) \sigma]^3 - \frac{1}{4} \tilde{c}_\sigma
  [g_{\sigma N}(n) \sigma]^4 - \frac{1}{4}\omega_{\mu\nu}
  \omega^{\mu\nu} + \frac{1}{2}m_\omega^2\omega_\mu \omega^\mu +
  \frac{1}{2}m_\rho^2 {\boldsymbol{\rho\,}}_\mu \cdot
       {\boldsymbol{\rho\,}}^\mu - \frac{1}{4}
       {\boldsymbol{\rho\,}}_{\mu\nu} \cdot
       {\boldsymbol{\rho\,}}^{\mu\nu} \, ,\end{array}
  \label{eq:Blag}
\end{myequation1}
where $g_{\sigma B}(n)$, $g_{\omega B}(n)$ and $g_{\rho B}(n)$ are
density dependent meson-baryon coupling constants and \mbox{$n =
\sum_{B=N,Y, \Delta} n_B$} is the total baryon number density. In~this
work we consider all members of the spin-$1/2$ baryon octet, which
include nucleons $N=(n,p)$ and hyperons $Y=(\Lambda, \Sigma^-,
\Sigma^0, \Sigma^+, \Xi^-, \Xi^0)$, as~well as the spin $3/2$ delta
isobar quartet $(\Delta^-, \Delta^0, \Delta^+$, $\Delta^{++})$.  The~density dependent coupling constants are given by~\cite{Typel:2018DD}
\begin{equation}
g_{i B}(n) = g_{i B}(n_0)\,\, a_{i}\,
\frac{1+b_{i}(\frac{n}{n_0}+d_{i})^{2}} {1+c_{i} (\frac{n}{n_0} +
  d_{i})^{2}}\, ,
\label{eq:gsigom}
\end{equation}
for $i=\sigma,\omega$, and~\begin{equation}
g_{\rho B}(n) = g_{\rho B}(n_0)\,\mathrm{exp}\left[\,-a_{\rho}
  \left(\frac{n}{n_0} - 1\right)\,\right]
\label{eq:grho}
\end{equation}
for the $\rho$ meson.  This approach is parametrized to reproduce the
properties of symmetric nuclear matter at saturation density $n_0$.
The parametrizations used in this work (GM1L, HV, and~DD2) are shown
in Table~\ref{table:parametrizations}. The~properties listed in this
table are the energy per nucleon ($E_0$), nuclear incompressibility
($K_0$), isospin asymmetry energy ($J$), and~effective nucleon mass
($m^*/m_N$).  The~isovector meson-baryon coupling constant is treated
density-dependent. Its numerical value has been fit to the slope of
the asymmetry energy ($L_0$) at $n_0$.
\begin{table}[H]
\centering
\caption{Properties of symmetric nuclear matter at saturation density
  for the hadronic parametrizations  {GM1L, DD2, and~HV} 
  considered in
  this work.}
\begin{tabular}{cccc}
\toprule \textbf{Saturation Property}& \textbf{GM1L}~
\cite{Spinella:2017thesis,Glendenning:1992pt} & \textbf{DD2} ~
\cite{Typel:2009a}  &\textbf{HV}~ \cite{Glendenning:1985a} \\
%\noalign{\smallskip}
\midrule
%\noalign{\smallskip}
$n_0$ (fm$^{-3}$) & 0.153 & 0.149    &0.145\\
$E_0$ (MeV)  & $-$16.30 & $-$16.02  &$-$15.98 \\
$K_0$ (MeV)  & 300.0 & 242.7        &285 \\
$m^*$/$m_N$  & 0.70 & 0.56          &0.77 \\
$J$ (MeV)  & 32.5 & 32.8            &36.8 \\
$L_0$ (MeV)  & 55.0 & 55.3          & --\\
\bottomrule
\end{tabular}
\label{table:parametrizations}
\end{table}
The parameters $a_{i}$, $b_{i}$, $c_{i}$, $d_{i}$, and~$a_\rho$ in
Equations~(\ref{eq:gsigom}) and (\ref{eq:grho}) are fixed by the binding
energies, charge~and diffraction radii, spin-orbit splittings, and~the
neutron skin thickness of finite nuclei.  The~density dependence of
the meson-baryon couplings of the DD2 parametrization eliminates the
need for non-linear self-interactions of the $\sigma$-meson field. The~non-linear terms~\cite{Boguta:1977a,Boguta:1983a} of the
$\sigma$-meson field in Equation \ (\ref{eq:Blag}) therefore only contribute
to the GM1L and HV models.  The~meson--hyperon coupling constants have
been determined following the Nijmegen extended soft core (ESC08)
model~\cite{Rijken:2010a}. The~relative isovector meson-hyperon
coupling constants were scaled with the hyperon isospin. For~the
$\Delta$-isobar we used $x_{{\sigma} {\Delta}} = x_{{\omega} {\Delta}}
= 1.1$ and $x_{{\rho} {\Delta}} = 1.0$, where $x_{ i \Delta} = g_{i
  \Delta}/g_{i N}$ (for details, see
Ref.~\cite{Spinella:2017thesis}).

The meson mean-field equations following from Equation \ (\ref{eq:Blag}) are
given by
\begin{eqnarray}
m_{\sigma}^2 \bar{\sigma} &=& \sum_{B} g_{\sigma B}(n) n_B^s -
\tilde{b}_{\sigma} \, m_N\,g_{\sigma N}(n) (g_{\sigma
  N}(n)\bar{\sigma})^2 - \tilde{c}_{\sigma} \, g_{\sigma N}(n) \,
(g_{\sigma N}(n) \bar{\sigma})^3 \, , \\ m_{\omega}^2
\bar{\omega} &=& \sum_{B} g_{\omega B}(n) n_{B}\, ,
\\ m_{\rho}^2\bar{\rho} &=& \sum_{B}g_{\rho B}(n)I_{3B} n_{B} \, ,
\end{eqnarray}
where $I_{3B}$ is the 3-component of isospin and $n_{B}^s$ and $n_{B}$
are the scalar and particle number densities of each baryon $B$, given
by
\begin{eqnarray}
n_{B}^s&=& \gamma_B \int \frac{d^3p}{(2 \pi)^3} \left[f_{B-}(p) -
  f_{B+}(p)\right] \frac{m_B^*}{E_B^*}, \\ n_{B}&=& \gamma_B \int
\frac{d^3p}{(2 \pi)^3} \left[f_{B-}(p) - f_{B+}(p)\right] \, .
\end{eqnarray}

The quantity $f_{B\mp}$ denotes the Fermi-Dirac distribution function,
and $E^*_B$ stands for the effective single-baryon energy,
\begin{equation}
f_{B\mp}(p)=\frac{1}{\exp\left[\frac{E_B^*(p) \mp \mu_B^*}{T}\right] +
  1},\,\,\,\, E_B^*(p)=\sqrt{p^2 + m_B^{*2}}\, . 
\end{equation}

The quantity $\gamma_B=2J_B+1$ accounts for the spin degeneracy, and~$m_B^*= m_B - g_{\sigma B}(n)\bar{\sigma}$ denotes the effective
baryon mass. The~effective baryon chemical potential, $\mu_B^*$, is
defined as
\begin{equation}
\mu_B^* = \mu_B - g_{\omega B}(n) \bar{\omega} - g_{\rho B}(n)
\bar{\rho} I_{3B} - \widetilde{R} \, ,
\end{equation}
where $\widetilde{R}$ is the so-called rearrangement term given by
\begin{eqnarray}
\widetilde{R} =\sum_B \left( \frac{\partial g_{\omega B}(n)}{\partial
  n} n_B \bar{\omega} + \frac{\partial g_{\rho B}(n)}{\partial n}
I_{3B} n_B \bar{\rho} - \frac{\partial g_{\sigma B}(n)}{\partial n}
n_B^s \bar{\sigma}\right) \, ,
\label{rear}
\end{eqnarray}
which is important for achieving thermodynamic consistency.  The~total
baryonic pressure of the matter follows from
\begin{eqnarray}
P &=& \sum_B \frac{\gamma_B}{3} \int \frac{d^3p}{(2 \pi)^3}
\frac{p^2}{E_B^*} [f_{B-}(p) + f_{B+}(p)] - \frac{1}{2} m_{\sigma}^2
\bar{\sigma}^2 + \frac{1}{2} m_{\omega}^2 \bar{\omega}^2 + \frac{1}{2}
m_{\rho}^2 \bar{\rho}^2  \nonumber \\ &-& \frac{1}{3} \tilde{b}_{\sigma} m_N
(g_{\sigma N}(n) \bar{\sigma})^3 - \frac{1}{4} \tilde{c}_{\sigma}
(g_{\sigma N}(n) \bar{\sigma})^4 + n \widetilde{R} \, , 
\label{HM:pressure}
\end{eqnarray}
while the energy density of the system follows from the standard Gibbs relation
$\epsilon(P,T, n) = - P + T S + \sum_B \mu_B n_B$.

To complete the description of the matter in the cores of (proto)
neutron stars, leptons ($L$)  need to be included in the
treatment. Leptons can be treated as free, relativistic Fermions whose
total thermodynamic potential is given by ($\gamma_L=2$)
\begin{equation}
\Omega_{\rm L} = - \sum_L \frac{\gamma_L}{3} \int \frac{d^3p}{(2 \pi)^3}
\frac{p^2}{E_L} [f_{L-}(p) + f_{L+}(p)] \, ,
\label{eq:leptons}
\end{equation}
with
\begin{eqnarray}
f_{L\mp}(p)&=&\frac{1}{\exp\left[\frac{E_L(p) \mp \mu_L}{T}\right] +
  1},\,\,\,\,\,\, E_L(p)=\sqrt{p^2 + m_L^{2}} \, .
\end{eqnarray}

The sum in Equation\ (\ref{eq:leptons}) is over electrons  and muons
 (with masses $m_L$) as well as over massless neutrinos,
$\nu_e$. Neutrinos do not contribute to cold neutron stars matter, but~are present in the cores of newly formed proto-neutron~stars.

\section{Electric Charge Neutrality, Chemical Equilibrium and Conserved~Charges}
\label{sec:conditions}

The composition of the matter in (proto-) neutron stars is constrained
by the conditions of electric charge neutrality, chemical equilibrium,
and baryon number conservation. Electric charge neutrality and baryon
number conservation leads to
\begin{equation}
\sum_B q_B\,n_B + \sum_L q_L\,n_L = 0  \,\,\,\,\,\, \textrm{and}
\,\,\,\,\,\, \sum_B \,n_B - n = 0 \, ,
\end{equation}
where $q_{B}$ and $q_{L}$ denote the electric charges of baryons and
leptons, respectively, and~$n_B$ and $n_L$ are the respective~number densities of these particles.  The~condition of chemical equilibrium
among the different particle species present at a given density reads
\begin{equation}
\mu_B = \mu_n + q_B (\mu_e - \mu_{\nu_e}) \, ,
\label{chem}
\end{equation}
where $\mu_n$, $\mu_e$, and~$\mu_{\nu_e}$ are the chemical potentials
of neutrons, electrons, and~neutrinos, respectively. The~lepton
chemical potentials obey
\begin{equation}
\mu_e = \mu_{\mu} + \mu_{\nu_e} + \mu_{\bar{\nu}_{\mu}} \, .
\end{equation}

Neutrinos are trapped in the core during the very early stages in the
life of a neutron star. Accepted lepton fractions at that stage of
stellar evolution are~\cite{Prakash:1997a}
\begin{eqnarray}
 Y_{Le} &=& \frac{n_e + n_{\nu_e}}{ n } = \xi ~ \simeq 0.4 \, ,
 \\ Y_{L\mu} &=& \frac{n_{\mu} + n_{\nu_{\mu}}}{ n } = 0 \, .
\end{eqnarray}

\begin{figure}[tbh]
\begin{center}
  \includegraphics[trim=0cm 2cm 0cm
    0cm,width=0.6\textwidth]{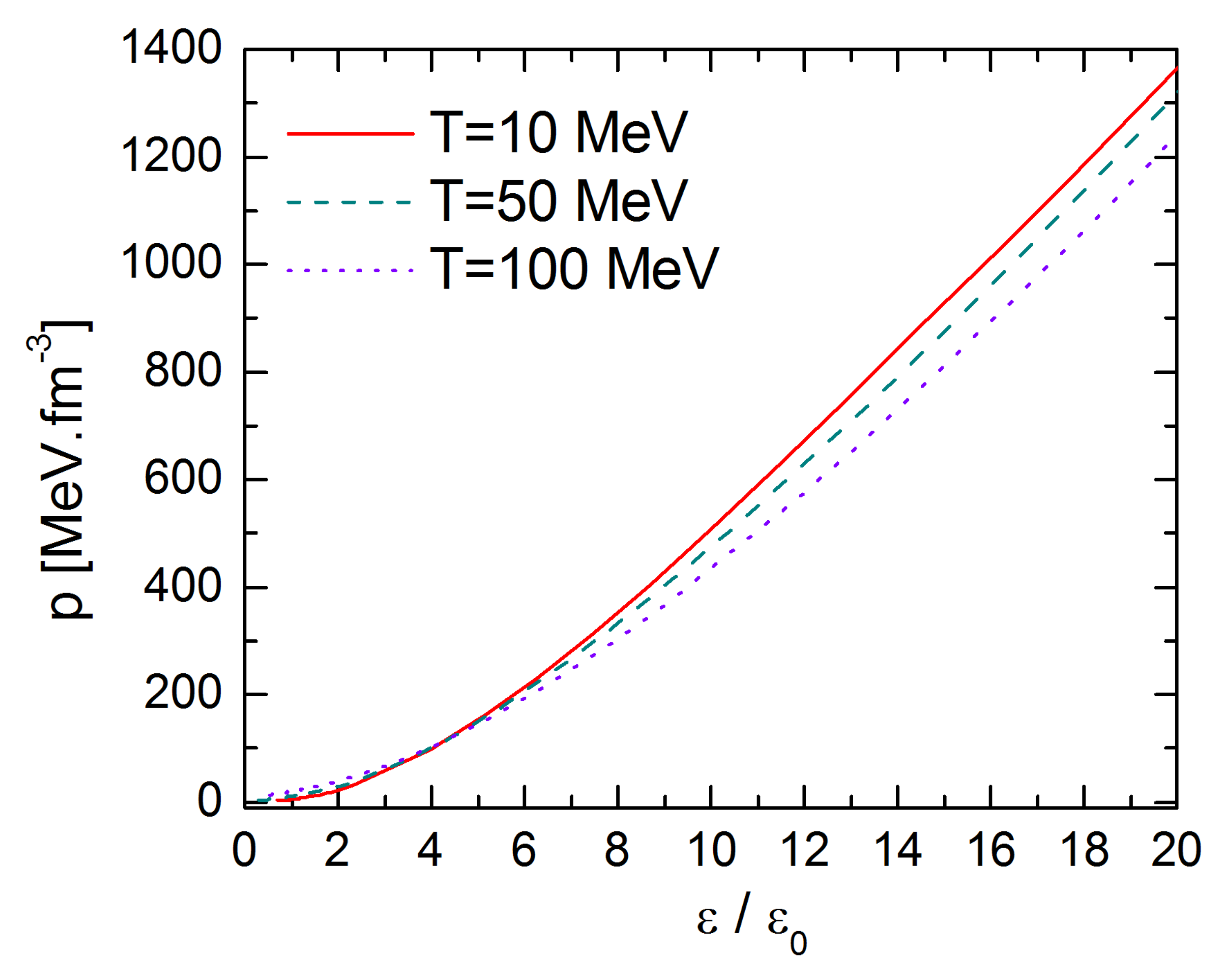}
\end{center}
\caption{Pressure versus energy density of hot neutron star matter
  computed for parameter set  {HV}.} 
\label{fig:eos}
\end{figure}

\begin{figure}[tbh]
\begin{center}
  \includegraphics[scale=0.25]{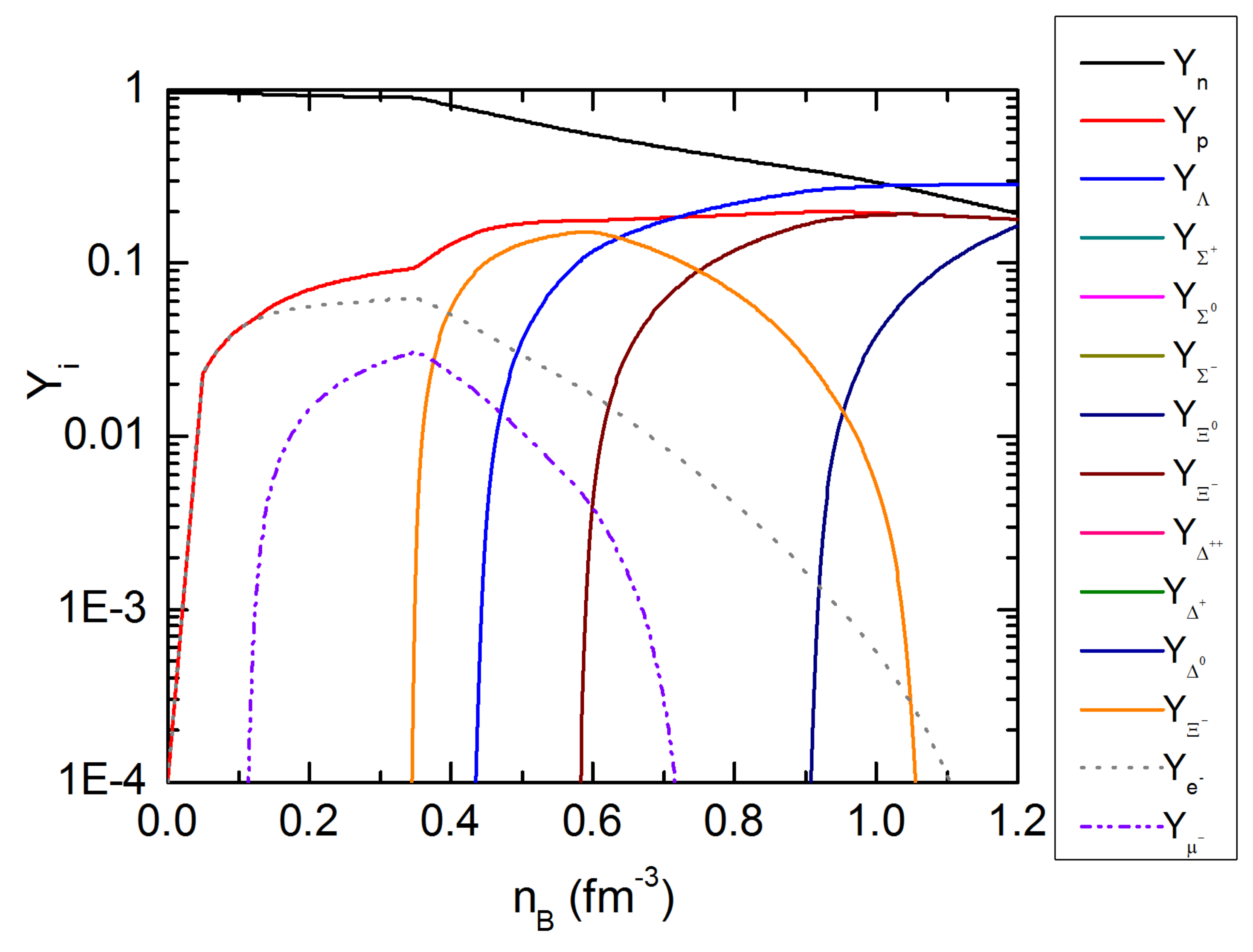}
      \includegraphics[scale=0.25]{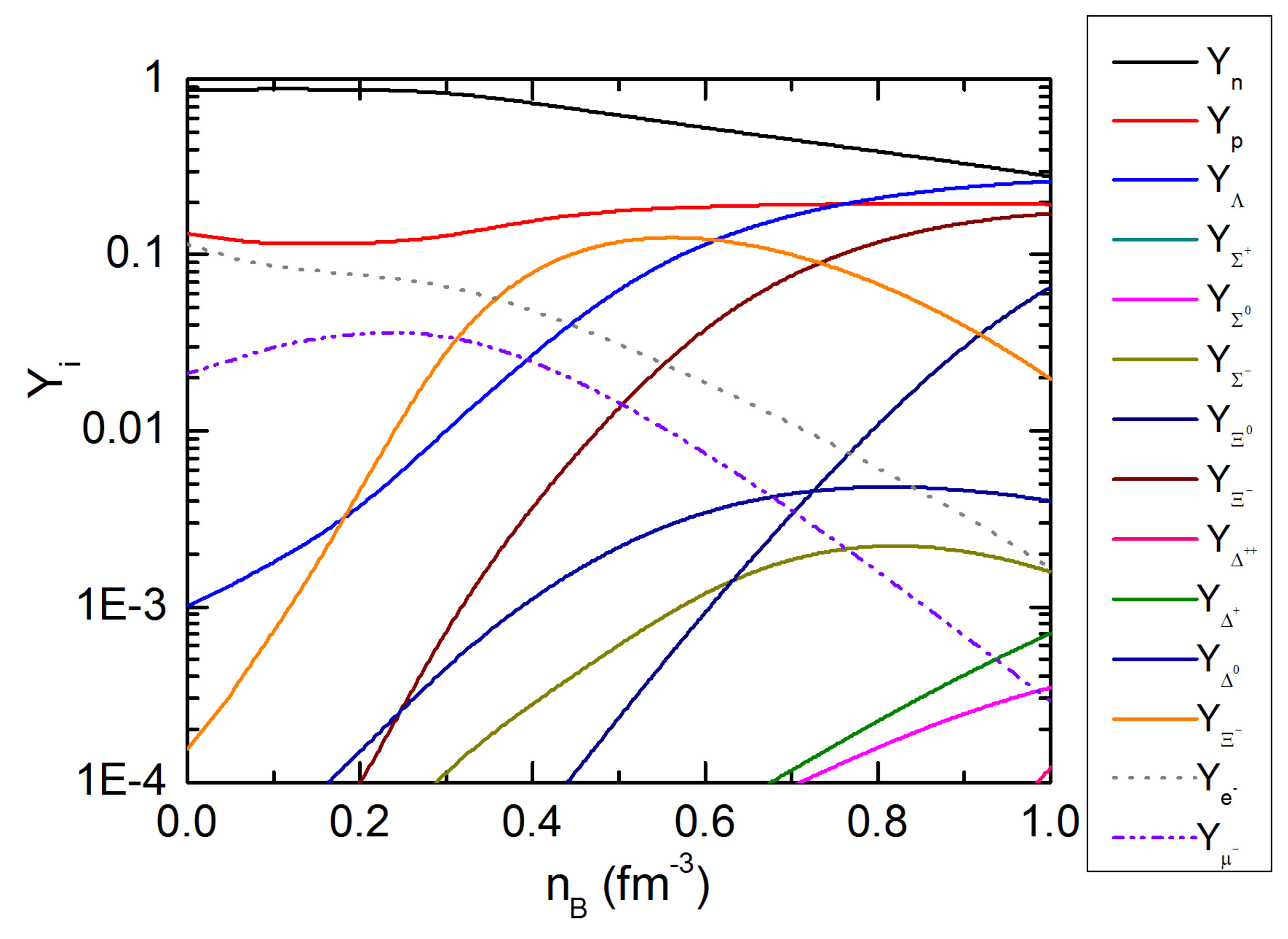}
\end{center}
\caption{Particle composition of neutron star matter at 0~MeV
  (\textbf{left}) and 25~MeV (\textbf{right}), computed for parameter
  set  {GM1L}. Neutrinos are not included.} 
  \label{fig:GM1Lpop}
\end{figure}

\begin{figure}[tbh]
\begin{center}
  \includegraphics[scale=0.25]{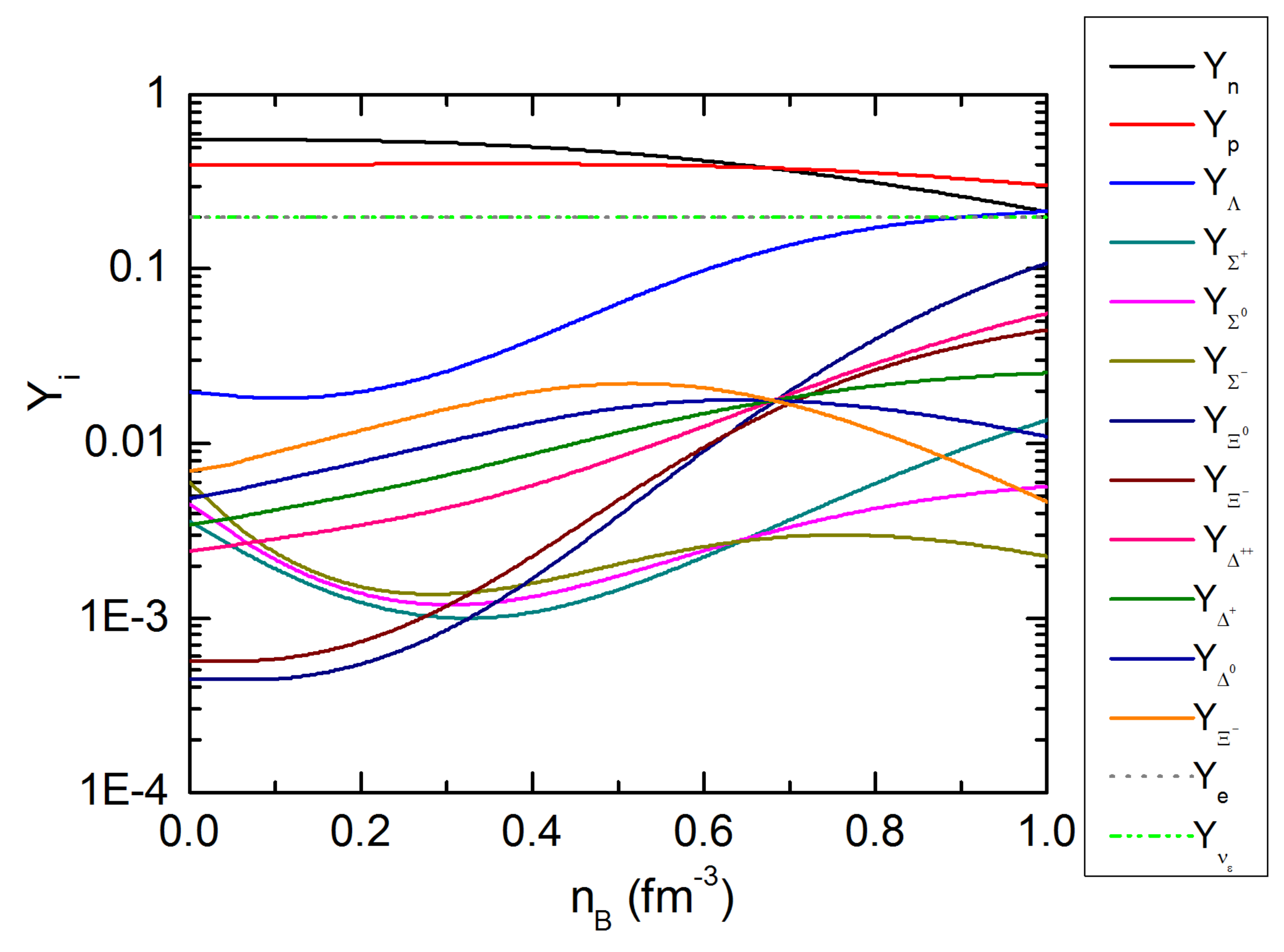}
    \includegraphics[scale=0.25]{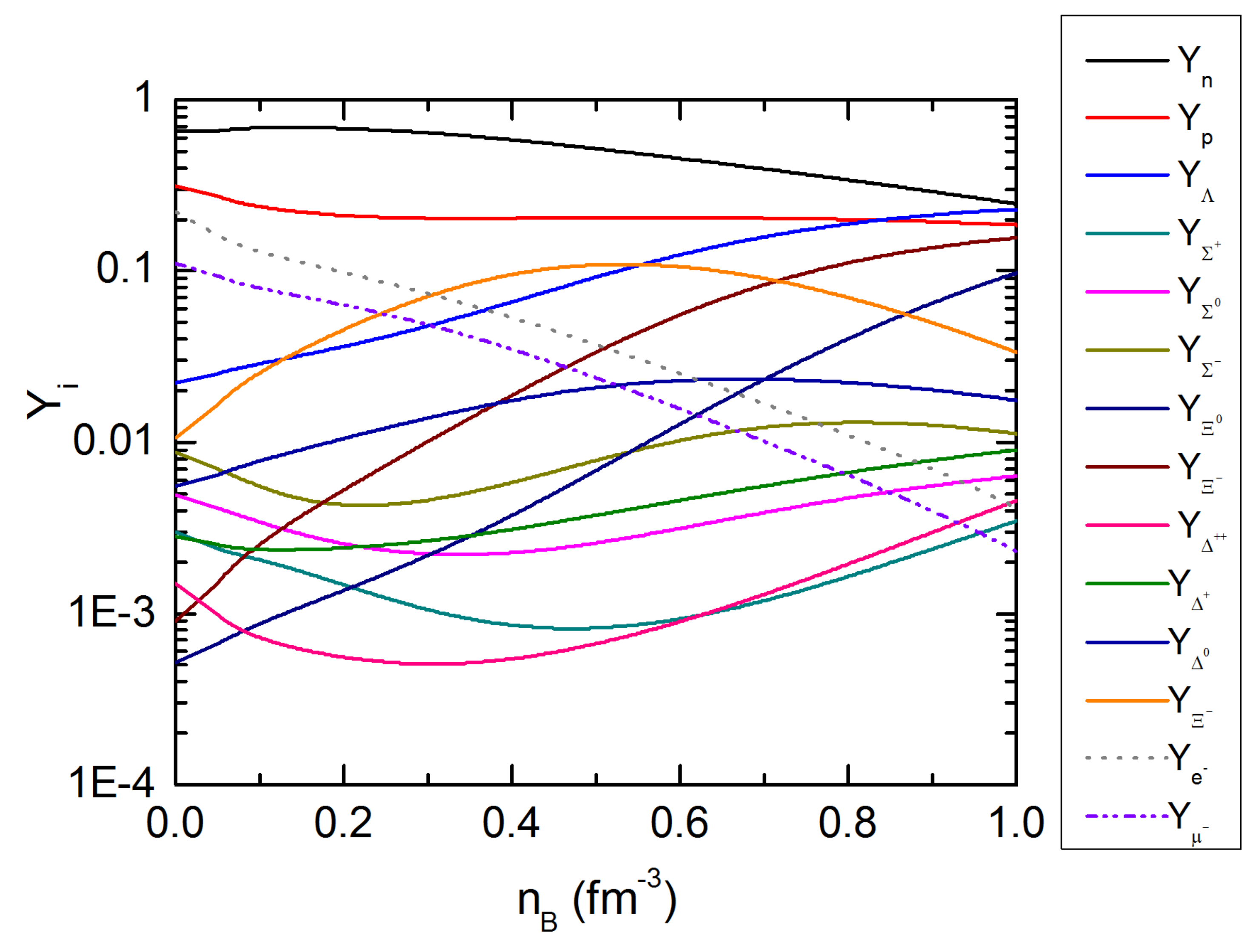}
\end{center}
\caption{Particle composition of neutron star matter at
  50~MeV with (\textbf{left}) and without (\textbf{right}) neutrinos, computed for
  parameter set GM1L.}
  \label{fig:GM1Lpop50}
\end{figure}

During this stage the system is characterized by three independent
chemical potentials, $\mu_n$, $\mu_e$, and~$\mu_{\nu_e}$, which,
together with the hadronic field equations and the conditions of
electric charge neutrality and baryon number conservation, determine
the composition of hot (proto-) neutron star matter. The~condition
$Y_{L\mu}=0$ has to be imposed in addition since no muons are present
when neutrinos are trapped.  When a proto-neutron star cools down,
the~matter in the core becomes transparent to neutrinos quickly,
leading to $\mu_{\nu_e} = \mu_{\bar{\nu}_{\mu}} = 0$ for cold neutron
stars. In~this case, the~number of independent chemical potentials
reduces from three to two ($\mu_n$ and $\mu_e$). Figure~\ref{fig:eos}
shows the equation of state (EoS) of neutron star matter at three
different temperatures. The~matter is composed of neutrons, protons,
hyperons, delta isobars, and~leptons in chemical equilibrium with each
other. Since the number of hyperons and delta isobars in the matter
increases drastically with temperature, the~pressure at a given
density drops for increasing~temperatures.
In Figures \ \ref{fig:GM1Lpop} and \ref{fig:GM1Lpop50} we show the
particle compositions of neutron stars at zero as well as finite
temperature.  As~can be seen, the~complexity of the compositions
intensifies quickly with increasing~temperature, lepton content,
and~assumptions about the neutrino~population.
\unskip
% Autor's reply: "(color online)" has been removed from this figure caption.

\section{Quark Matter Modeled with the Non-Local NJL~Model}
\label{sec:quark_eos_nonlocalNJL}

A model widely used to describe the properties of deconfined 3-flavor
quark matter is the Nambu--Jona-Lasinio
model~\cite{Buballa:2005a}. In~the following, we present a non-local
extension of this model (denoted n3NJL). The~effective action of this
model reads (for details, see
Refs.\ \cite{Orsaria:2013a,Orsaria:2014a})
\begin{myequation2}
\begin{array}{ccl}
S_E &=& \int d^4x \left\{ \bar \psi (x) \left[ -i
  \partial{\hskip-2.0mm}/ + \hat m \right] \psi(x) - \frac{G_S}{2}
\left[ j_a^S(x) \ j_a^S(x) + j_a^P(x) \ j_a^P(x) \right] \right.
 \\ 
& & \qquad \qquad \left. - \frac{H}{4} \ T_{abc} \left[
  j_a^S(x) j_b^S(x) j_c^S(x) - 3\ j_a^S(x) j_b^P(x) j_c^P(x)
  \right]\right.  \left.- \frac{G_{V}}{2} \left[j_{V}^\mu(x)
  j_{{V}, \mu}(x)\right]\right\}\, ,\end{array}
  \label{L3}
\end{myequation2}
where $\psi \equiv (u, d, s)^T$ and $\hat m$ denotes the current quark
mass matrix give by $\hat m = {\rm diag}(m_u, m_d, m_s)$. The~currents
$j_a^{S,P}(x)$ and $j_{V}^{\mu}(x)$ depend on form factors which
describe non-local interactions among the quarks. The~quantity
$T_{abc}$ in the 't Hooft term accounts for quark-flavor mixing.
The~current quark mass $\bar{m}$ of up and down quarks and the
coupling constants $G_S$ and $H$ in Equation\ (\ref{L3}) are fitted to
the pion decay constant, $f_\pi$, and~meson masses $m_{\pi}$,
 {$m_K$}, $m_{\eta'}$ \cite{Contrera:2008a,Contrera:2010a}. This
leads to $\bar{m} = 6.2$~MeV, $\Lambda = 706.0$~MeV, $G_S \Lambda^2 =
15.04$, and $H \Lambda^5 = - 337.71$. The~current mass of the strange
quark is treated as a free parameter. Its value is set to $m_s
=140.7$~MeV.  It is customary to express the vector interaction,
$G_V$, in~terms of the strong coupling constant, $G_S$, according to
$\xi_V \equiv G_V/G_S$. The~value of $\xi_V$ is then typically varied
from $0$ to around $0.09$.

The expression of the thermodynamic potential that follows from 
 $S_E$ of Equation\ (\ref{L3}) is given, for~the mean-field approximation, by~
\begin{myequation3}
\begin{array}{lll}
    &&\Omega = -\frac{3}{\pi^3}\sum_{f=u,d,s}
    \int_0^{\infty}dp_0 \int_0^{\infty} dp \ \mathrm{ln}
    \Biggl\{ \left[ \widehat{\omega}^2_f+M_f^2(\omega^2_f)\right]
    \frac{1}{\omega^2_f+m^2_f}\Biggr\} 
    -\frac{3}{\pi^2}\sum_{f=u,d,s}\int_0^{\sqrt{\mu^2_f-m^2_f}}
    dp\,p^2 \times \\
    &&\left[\left(\mu_f-E_f\right)\theta\left(\mu_f-m_f\right)\right]
    \quad -\frac{1}{2} \Bigg[\sum_{f=u,d,s}\left(\bar{\sigma}_f\bar{S}_f+
    \frac{G_S}{2}\bar{S}^2_f\right)+\frac{H}{2}\bar{S}_u\;\bar{S}_d\;\bar{S}_s
    \Biggr] 
   - \sum_{f=u,d,s}\frac{\overline{\omega}_f^2}{4G_V} \, . \end{array} \label{grandpotential}
\end{myequation3}

The quantities $\bar{\sigma}_f$, $\overline{\omega}_f$, and~$\bar{S}_f$ denote the scalar, vector, and~auxiliary mean-fields of
quarks, respectively.  The~quantity $E_f$ denotes the energy-momentum
relation of free quarks given by $E_f= \sqrt{{\boldsymbol p}^2 +
  m^2_f}$.  The~symbol $\omega^2_f$ \,stands for $\omega^2_f =
(p_0+i\mu_f)^2 + {\boldsymbol p}^2$. The~relation of the
momentum-dependent quark masses reads $M_f(\omega_f^2) =
m_f+\bar{\sigma}_f g(\omega^2_f)$.  The~quantities $g(\omega^2_f) =
\mathrm{exp}(-\omega^2_f/\Lambda^2)$ are Gaussian form factors, which
describe the nonlocal nature of the interactions among quarks
~\cite{Orsaria:2014a}.  The~quantities $\bar{S}_f$ denote auxiliary
mean fields~\cite{Scarpettini:2004a}.  The~scalar and vector quark
mean-fields are obtained by minimizing the thermodynamic potential,
$\partial \Omega / \partial \bar{\sigma}_f = 0$ and $\partial \Omega /
\partial \overline{\omega}_f = 0$. The~number densities of quarks are
obtained from the thermodynamic potential as $n_f = \partial \Omega /
\partial \mu_f$.

To determine the equation of state of the system is given in terms
of  the quark mean-fields $\bar{\sigma}_f$, $\bar{\omega}_f$ and
the neutron and electron chemical potentials $\mu_n$ and $\mu_e$.  The~pressure $p_Q$ and the energy density $\epsilon_Q$ of the system at
zero temperature are given by
\begin{equation} \label{eq:qpressure}
  P_Q = \Omega_0-\Omega \, ,
\end{equation}
and
\begin{equation} \label{eq:qdensity}
  \epsilon_Q = -P_Q + \sum_{f=u,d,s} n_f \mu_f +
  \sum_{\lambda=e^-,\mu^-} n_{\lambda} \mu_{\lambda} \, ,
\end{equation}
where $\Omega_0$ is chosen such that $P_Q$ vanishes at $T=\mu=0$
\cite{Orsaria:2013a,Orsaria:2014a}.

\section{Pasta Structure of the Hadron-Quark~Phase}
\label{sec:structure}

In this section, we provide a brief discussion of the mathematical
treatment of the mixed phase region of hadrons and quarks in neutron
stars. The~starting point is the Gibbs condition for phase equilibrium
between hadronic ($H$) matter and quark ($Q$) matter
~\cite{Glendenning:1992pt,Glendenning:2001pr},
\begin{eqnarray}
  P_H ( \mu_n , \mu_e, \{ \phi_H \} ) = P_Q (\mu_n , \mu_e, \{ \phi_Q
  \}) \, ,
\label{eq:gibbs}
\end{eqnarray}
where $P_H$ and $P_Q$ denote the pressures of hadronic matter and
quark matter, respectively
~\cite{Glendenning:1992pt,Glendenning:2001pr}. The~quantities $\{
\phi_H \}$ and $\{ \phi_Q \}$ in Equation\ (\ref{eq:gibbs}) collectively
denote the field variables of hadronic ($\bar{\sigma}$, $\bar\omega$,
$\bar\rho$) and quark ($\bar\sigma_f$, $\bar\omega_f$, $\bar S_f$)
matter as well as the Fermi momenta ($k_B$, $k_\lambda$) that
characterize a solution of the equations of hadronic matter described
in Section~\ref{sec:nucl_eos}. The~quantity $\chi \equiv V_Q/V$
describes the volume proportion of quark matter, $V_Q$, in~the unknown
volume $V$.  $\chi$, therefore, varies between 0 and 1, where~0 means
that 0\% quark matter is present and 1 means that the matter consist
of 100\% quark matter.  Equation~(\ref{eq:gibbs}) is solved subject to
the conditions of global baryon number conservation and global
electric charge conservation.  The~global conservation of baryon
charge is expressed as~\cite{Glendenning:1992pt,Glendenning:2001pr}
\begin{eqnarray}
  n_b = \chi \, n_Q(\mu_n, \mu_e, \{ \phi_Q \} ) + (1-\chi) \, n_H
  (\mu_n, \mu_e, \{ \phi \}) \, ,
\label{eq:mixed_rho}
\end{eqnarray}
where $n_Q$ and $n_H$ are the baryon number densities of the quark and
hadronic phases, respectively. Global electric charge neutrality
is given by~\cite{Glendenning:1992pt,Glendenning:2001pr}
\begin{eqnarray}
  0 = \chi \ n_Q(\mu_n, \mu_e, \{ \phi_Q \} ) + (1-\chi) \ n_H
  (\mu_n, \mu_e, \{ \phi \}) \, ,
\label{eq:mixed_charge}
\end{eqnarray}
\textls[-15]{where $q_Q$ and $q_H$ denoting the electric charge densities of the
quark phase and hadronic phase, respectively.}

As demonstrated by Glendenning
~\cite{Glendenning:1992pt,Glendenning:2001pr}, a~mixed phase of
hadron-quark matter may arrange itself in such a way that the total
energy of the system is minimized. For~matter described by global
electric charge neutrality, this is the same as minimizing the
contributions to the total energy due to phase segregation, which
includes the surface and Coulomb energy contributions. Mathematically,
the~relations for the Coulomb ($\epsilon_C$) and the surface
($\epsilon_S$) energy densities are given by
~\cite{Glendenning:1992pt,Glendenning:2001pr,Spinella:2018U}
\begin{eqnarray}
  \mathcal{E}_C &=& 2\pi e^2 \left[ q_H(\chi) - q_Q(\chi)
    \right]^2 r^2 x f_D(x) \, ,
    \label{eq:eps_c} \\
    \mathcal{E}_S &=& D x \alpha(\chi)/r\, ,
    \label{eq:eps_s}
\end{eqnarray}
where $q_H$ denotes the charge density of the hadronic phase and $q_Q$
stands for the charge density of the quark phase. The~symbol $r$ is
the radius of the rare phase structures (i.e., quark blobs embedded in
hadronic matter at the onset of phase equilibrium).  The~surface
tension between hadronic matter and quark matter is denoted by
$\alpha(\chi)$ where $0 \leq \chi \leq 1$.  The~quantity $f_D(x)$ in
Equation\ (\ref{eq:eps_c}) is defined as
\begin{equation}
  f_D(x) = \frac{1}{D+2} \left[ \frac{1}{D-2} (2-D\, x^{1-2/D}) + x
    \right]\,  ,
  \end{equation}
where $D$ is the dimensionality of the lattice and $x \equiv
\mathrm{min}(\chi,1-\chi)$.  The~phase rearrangements will result in
the formation of geometrical structures of the rare phase distributed
in a lattice that is immersed in the dominant phase.  For~the sake of simplicity, the~rare phase structures are approximated by
spherical blobs, rods, and~slabs. The~spherical blobs populate sites
in a three dimensional ($D=3$) body centered cubic (BCC) lattice, the~rods in a two dimensional ($D=2$) triangular lattice, and~the slabs in
a simple one dimensional ($D=1$) lattice. At~a volume ratio of $\chi =
0.5$ both hadronic and quark matter exist as slabs in the same
proportion, and~at $\chi > 0.5$ the hadronic phase becomes the rare
phase with its geometry evolving in reverse order (from slabs to rods
to blobs) with increasing density~\cite{Glendenning:1992pt,Glendenning:2001pr,Spinella:2018U}.

The determination of the surface tension, $\alpha$, of~the
quark-hadro interface is hampered by the fact that a single theory
that can accurately describe both hadronic matter and quark matter
does not exist. We therefore make use of an approximation proposed by
Gibbs where the surface tension is taken to be proportional to the
difference in the energy densities of the interacting phases,
\begin{equation}
  \alpha(\chi)=\eta L
  \left[\mathcal{E}_Q(\chi)-\mathcal{E}_H(\chi)\right]\, .
\end{equation}

\begin{figure}[tbh]
\centering
\includegraphics[trim=0cm 0.5cm 0cm 0cm,width=0.9\textwidth]{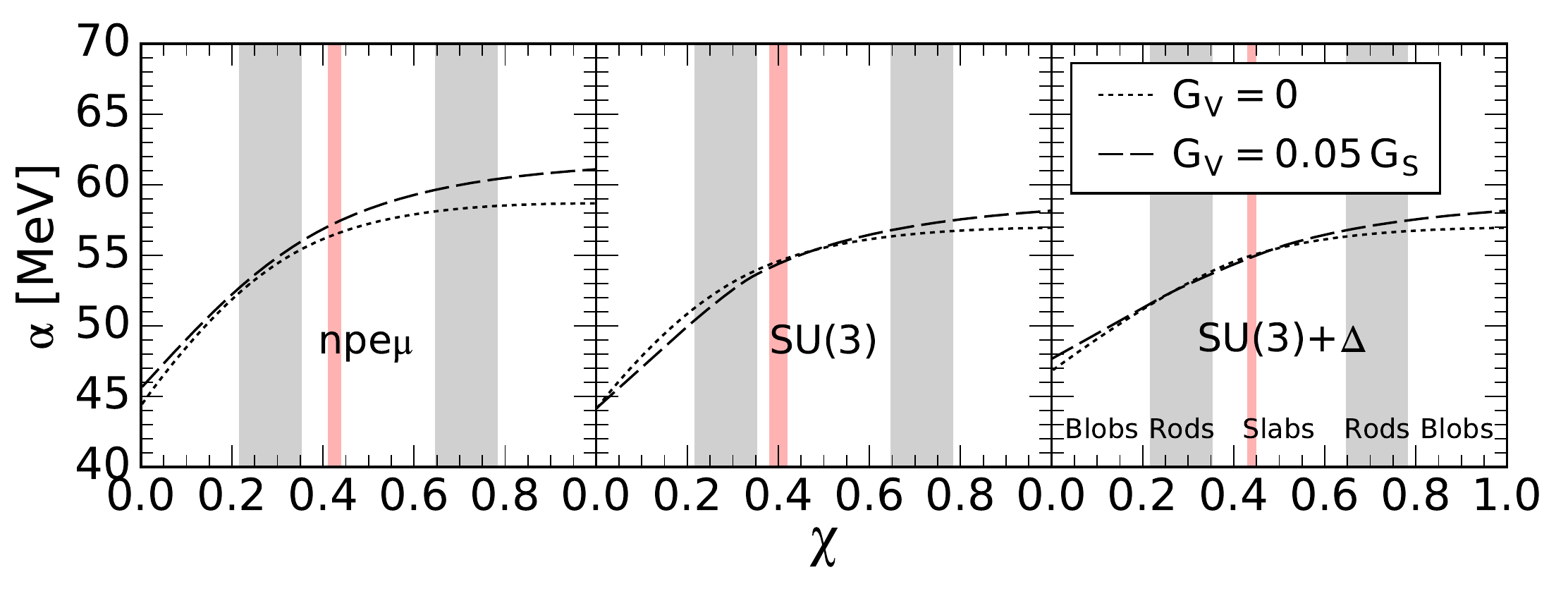}% \\ %\vspace{0.5cm}
\caption{Surface tension $\alpha$ in the hadron-quark mixed phase for
  the GM1L parameter set~\cite{Spinella:2017thesis}.}
\label{fig:st.GM1L}
\end{figure}

The quantity $L$ is proportional to the surface thickness which should
be on the order of the range of the strong interaction (1
fm). The~quantity $\eta$ denotes a proportionality constant. For~the
results shown in this work we maintained the energy density
proportionality but set the~parameter $\eta = 0.08$ so that the
surface tension falls below \mbox{70~MeV ~fm$^{-2}$}, which is a
reasonable upper limit for the existence of a hadron-quark mixed
phase~\cite{Yasutake:2014a}. The~surface tension as a function of
$\chi$ is given in Figure~\ref{fig:st.GM1L} for the nuclear GM1L
parametrization introduced in Section~\ref{sec:nucl_eos}.

The size of the rare phase structures is given by the radius ($r$) and
is determined by minimizing the sum of the Coulomb and surface
energies, $\partial(\mathcal{E}_C+\mathcal{E}_S) / \partial r$, and~solving for $r$
\begin{equation}
  r(\chi) = \left(\frac{D\alpha(\chi)}
    {4\pi e^2f_D(\chi)\left[q_H(\chi)
  - q_Q(\chi)\right]^2}\right)^\frac{1}{3}.
\end{equation}

Rare phase structures are centered in the primitive cell of the
lattice, taken to be a Wigner-Seitz cell of the same geometry as the
rare phase structure. The~Wigner--Seitz cell radius $R$ is set so that
the primitive cell is charge neutral, i.e.,~$R(\chi) = rx^{-1/D}$.

\begin{figure}[tbh]
\centering \includegraphics[trim=0cm 0.5cm 0cm 0cm,width=0.9\textwidth]{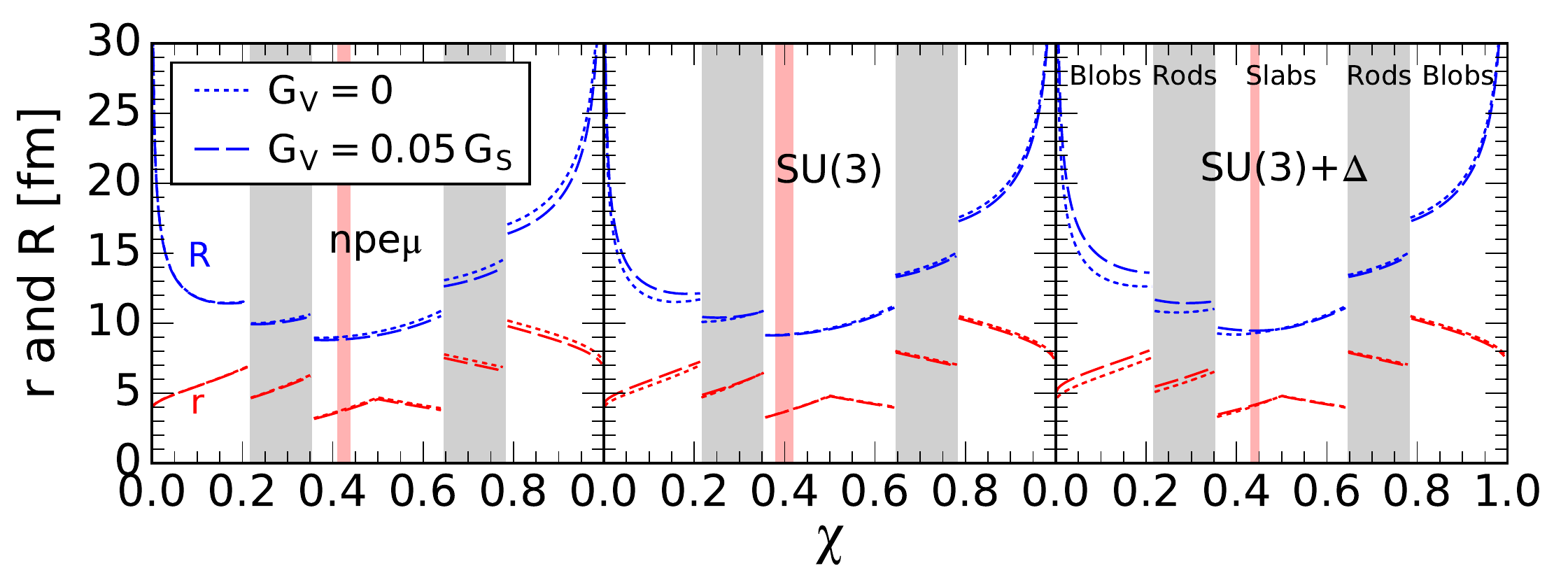}%\\ %\vspace{0.5cm}
\caption{Radius of the rare phase structure, $r$, and~of the
  Wigner-Seitz cell, $R$, in~the quark-hadro mixed phase computed
  for the DD2 parametrization~\cite{Spinella:2017thesis}. A~similar
  figure for the GM1L parametrization can be found in
  Ref.~\cite{Spinella:2017thesis}.}
\label{fig:structure-radii}
\end{figure}

In Figure~\ref{fig:structure-radii} we show $r$ and $R$ as a function of the quark
fraction in the mixed phase. Both $r$ and $R$ 
increase with an increase in the baryonic degrees of freedom, particularly when
$\chi \lesssim 0.5$ and the vector interaction is~included.
The transport properties of a hypothetical quark-hadro Coulomb
lattice in the core of a neutron star and the neutrino emissivity in
this phase have been investigated in
Refs.~\cite{Na:2012a,Spinella:2016EJP,Spinella:2018U,Freeman:2019AN}.

\section{Hadron-Quark Lattices in the Cores of Rotating Neutron~Stars}\label{sec:GR}

In this section we use the equations of state discussed in
Sections~\ref{sec:nucl_eos} and \ref{sec:quark_eos_nonlocalNJL} to
explore the possible existence of hadron-quark lattices in the cores
of (cold) rotating neutron stars. Depending on rotational frequency,
such stars are highly deformed and experience significant density
changes (up to 50\%) in their cores which drastically alters their
core compositions.  The~line element that is used to model rotating
neutron stars has the form~\cite{Friedman:1986a}
\begin{eqnarray}
  d s^2 = - e^{2\nu} dt^2 + e^{2\psi} (d\phi - \omega dt)^2 + e^{2\mu}
  d\theta^2 + e^{2\lambda} dr^2 \, ,
\label{eq:metric} 
\end{eqnarray} where   $\nu$, $\psi$,  $\mu$ and
$\lambda$ denote metric functions and $\omega$ stand for the angular
velocity of the local inertial frames. Each quantity depends on the
radial coordinate $r$, the~polar angle $\theta$, and~on the rotating
star's angular velocity $\Omega$.  The~metric functions and the frame
dragging frequency are to be computed from Einstein's field equation,
\begin{equation} 
R^{\kappa\sigma} - \frac{1}{2} R g^{\kappa\sigma} = 8 \pi
T^{\kappa\sigma} \, ,
\label{eq:einstein}
\end{equation}
where $T^{\kappa\sigma} = T^{\kappa\sigma}(\epsilon, P(\epsilon))$
denotes the energy momentum tensor of the stellar matter, whose
equation of state is given by $P(\epsilon)$. The~other quantities in
Equation\ (\ref{eq:einstein}) are the Ricci tensor $R^{\kappa\sigma}$, the~curvature scalar $R$, and~the metric tensor, $g^{\kappa\sigma}$. A~strict limit on rapid rotation is given by the Kepler frequency,
${\Omega_{\,\rm K}}$, at~which mass shedding from the equator
terminates stable rotation. This frequency is of great theoretical
value because it sets an absolute limit on rapid rotation, but~it may
not be accessible to neutron stars as other instabilities set tighter
constraints on stable rotation~\cite{Weber:1999book,Patruno:2012}. As~a case in point, even the most rapidly spinning neutron star observed
to date, PSR J1748-244ad, spinning at 716 Hz (i.e., 1.4 milliseconds)
\cite{Hessels:2006}, rotates only at a fraction of the Kepler
frequency.

The general relativistic expression for the Kepler frequency is
obtained by evaluating $\delta \int ds^2 =0$ in the star's equatorial 
plane, which leads to~\cite{Weber:1999book,Friedman:1986a}
\begin{eqnarray}
  {\Omega_{\,\rm K}} = \omega +\frac{\omega_{,r}} {2\psi_{,r}} +
  e^{\nu -\psi} \sqrt{ \frac{\nu_{,r}} {\psi_{,r}} +
    \Bigl(\frac{\omega_{,r}}{2 \psi_{,r}} e^{\psi-\nu}\Bigr)^2 } \, ,
\label{eq:okgr}
\end{eqnarray}
where ${}_r \equiv \partial/\partial r$. This equation is to be
computed self-consistently in combination with Einstein's field
equation at the equator of a rotating neutron~star.

\begin{figure}[tbh]
\centering
\includegraphics[trim=0cm 0.8cm 0cm 0.7cm,width=0.6\textwidth]{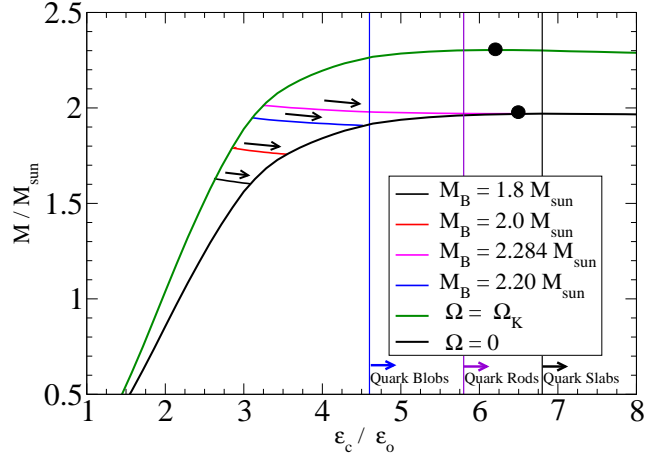}% \\ %\vspace{.5cm}
\caption{(color online) Gravitational mass as a function of central
  stellar density (in units of the energy density at nuclear
  saturation, $\epsilon_0=140$ MeV/fm$^3$) of non-rotating and
  rotating neutron stars for nuclear equation of state (EoS) GM1L. Shown are several
  stellar paths that would be followed by neutron stars with a
  constant baryon mass, $M_B$, as~they spin down from their respective
  Kepler frequencies (curve labeled $\Omega=\Omega_{\, \rm K}$ to zero
  frequency (curve labeled $\Omega =0$).}
  \label{fig:MvsecA}
\end{figure}

In Figure~\ref{fig:MvsecA} we show the gravitational mass versus central
energy density relationships for both neutron stars spinning at their
Kepler frequencies, as~well as non-rotating neutron stars. As~one moves
along either of these two lines, the~baryon mass $M_B$ of the stars
changes, increasing with larger central density. Isolated neutron
stars spinning down due to the loss of energy would follow a path of
constant baryon mass, from~the Kepler frequency curve down to the
non-rotating one. Four such evolutionary paths (marked with arrows)
are depicted in Figure \ \ref{fig:MvsecA}.
\vspace{-6pt}
One sees that neutron stars with baryon masses less than $2.20\,
M_\odot$ do not become dense enough during spin down to produce
hadron-quark phases in their cores. This is only possible for the most
massive neutron stars of our study, which contains a Coulomb lattice
made of quark blobs and quark rods if the star's rotational frequency
has dropped to 300 Hz (25\% of $\Omega_{\, \rm K}$). Other quark phases
(\mbox{e.g., quark slabs}) would not be present in stable neutron~stars.

\section{Quark Matter at Finite Temperature Modeled with the Local PNJL~Model}
\label{sec:quark_eos_localNJL}

The non-local 3-flavor NJL model has been introduced in
Section~\ref{sec:quark_eos_nonlocalNJL}. This model has then be used,
together with the model for hadronic matter introduced in
Section~\ref{sec:nucl_eos}, to~determine the possible possible pasta
structure of hadrons and quarks in the hadron-quark mixed phase of
neutron star matter.  In~this section, we introduce a local version of
the  {Polyakov--Nambu--Jona-Lasinio}
% Autor's comment:  Polyakov--Nambu--Jona-Lasinio has been added. Note: use only single dash
% for Jona-Lasinio
(PNJL) model which will be used to study quark matter at zero as well
as at finite temperatures
~\cite{Malfatti:2019,Shao:2011a,Lenzi:2012a,Bonnano:2012a}.
 {The~Polyakov loop effective potential, which relates quark
  confinement to color confinement, is included, so that from here on
  we refer to this model as PNJL.}  The~general expression for the
mean-field thermodynamic potential of the model can be written as
\begin{equation}
\Omega = \Omega_{\rm reg} + \Omega_{\rm free} + \Omega_0 + \Omega_{\mbox{v}}
+\mathcal{U}(\Phi ,T)\, ,
\label{eq:Omega}
\end{equation}
where $\Omega_0$ is defined by the condition that $\Omega$ vanishes at
zero temperature and chemical potential, $T=\mu=0$.  Allowing for
three independent chemical potentials $\mu_f$ for the three quark
flavors $f=(u, d, s)$, the~SU(3) regularized thermodynamic
potential in the presence of the quark condensates $\sigma_f =
\left\langle\bar{f}f\right\rangle$ reads
\begin{eqnarray}
\Omega_{\rm reg} = &-& 2
N_c\sum_{f}\int_0^{\Lambda}\frac{\mathrm{d}^3p}
{\left(2\pi\right)^3}\,{E_f} + G_s (\sigma_u^2 + \sigma_d^2 +
\sigma_s^2) + 4 H \sigma_u \sigma_d \sigma_s \, , \label{omegaT_njl}
\end{eqnarray}
and the corresponding free term is given by\vspace{12pt}
\begin{eqnarray}
\Omega_{\rm free} = &-&2T \sum_{f, c}\int
_0^{\infty}\,\frac{\mathrm{d}^3p} {\left(2\pi\right)^3}\,\Bigg\{
      {\mathrm{ln} \left(1+ e^{-\frac{E_f - \mu_f + i
            c\phi_3}{T}}\right)} + {\mathrm{ln} \left(1+ e^{-\frac{E_f
            + \mu_f - i c\phi_3}{T}}\right) \Bigg\} } \,
      , \label{omegaT_free}
\end{eqnarray}
where $N_c = 3$ is the number of quark colors and $E_{f} =
\sqrt{p^{2}+M_{f}^{2}}$ is the energy of a quark of flavor $f$. The~color background fields due the coupling to the Polyakov loop are $
\phi_c = c \phi_3$, i.e.,~$\phi_r = - \phi_g = \phi_3$ and $\phi_b =
0$. The~sums over the flavor and color indices are over $f = (u, d, s)$ and
$c = ( r, g, b)$, respectively. The~constituent quark masses $M_{f}$
are given by
\begin{equation}
M_{f}=m_{f}-2G_s \sigma_f - 2 H \sigma_j \sigma_k \, , 
\end{equation}
with $f,j,k=u,d,s$ indicating cyclic permutations of the quark
flavors. For~our calculations we have used the Polyakov loop effective
potential as given in Ref.~\cite{Roessner:2006a},
\begin{eqnarray}
{\cal{U}}(\Phi ,T) &=& \left[-\,\frac{1}{2}\, a_1(T, T_0)\,\Phi^2 +
  a_2(T, T_0)\, \ln(1 - 6\, \Phi^2 + 8\, \Phi^3 - 3\, \Phi^4)\right]
T^4 \, ,
\label{effpot}
\end{eqnarray}
where the coefficients $a_1(T, T_0)$ and $a_2(T, T_0)$ are fixed by
Lattice QCD simulations of gluon dynamics and the traced Polyakov loop
$\Phi = [ 2 \cos(\phi_3/T) + 1]/3$ \cite{Contrera:2010a,Roessner:2007a}.
The parameter $T_0$ denotes the critical temperature of the
deconfinement phase transition.  The~effective potential is key to
describe the phase transition from the color confined state ($T <
T_0$, with~the minimum of the effective potential at $\Phi = 1$) to
the color deconfined state ($T > T_0$, with~the minima of~the
effective potential at $\Phi = 0$).  $T_0$ is the only free parameter
of the Polyakov loop once the effective potential has been chosen. In~this work we will set $T_0=195.0$ MeV~\cite{Carlomagno:2013a}.  The~scalar coupling constant, $G_s$, the~'t Hooft coupling constant, $H$,
the~quark masses, and~the three-momentum ultraviolet cutoff,
$\Lambda$, are model parameters. The~respective values of these
quantities are $m_u=m_d=5.5$ MeV, $m_s=140.7$ MeV, $\Lambda=602.3$
MeV, $G_s\Lambda^{2}=3.67$ and $H\Lambda^{5}=-12.36$
\cite{Rehberg:1996a}.  The~grand canonical thermodynamic potential
$\Omega_{\mbox{v}}$ due to vector interactions among quarks is given
by
\begin{equation}
\Omega_{\mbox{v}} = - G_{\mbox{v}}\, \sum_f n_f^2 \, ,
\end{equation}
where the number density of a quark of flavor $f$ in the mean field
approximation is given by
\begin{equation}
n_f=\frac{N_c}{3\pi^2}[(\mu_f - 2G_{\mbox{v}} n_f)^2-M_f^2]^{3/2} \, .
\end{equation}

As already mentioned in Section~\ref{sec:quark_eos_nonlocalNJL}, the~vector coupling constant $G_{\mbox{v}}$ is treated as a free parameter
and is generally expressed in terms of the strong coupling constant
$G_s$, that is, $\xi_{\mbox{v}} \equiv G_v/G_s$. 

The minimization of the thermodynamic potential with respect to the
quark condensates and the Polyakov-loop color field $\phi_3$ leads to
a system of coupled, non-linear equations that are to be solved
numerically together with the conditions of electric charge
neutrality, baryon number conservation, and~chemical equilibrium,
similar to the ones discussed in Section~\ref{sec:conditions} for
hadronic matter.  Once the thermodynamic potential has been computed,
the system's pressure, $P=-\Omega$, and~the quark number density,
$n_q = \sum_f n_f$, can readily be determined. The~EoS of quark matter
then follows as $\epsilon (P,T, n_f) = - P + T S + \sum_f \mu_f n_f$,
where $S = \frac{\partial P}{\partial T}$ and $n_f = \frac{\partial
  P}{\partial \mu_f}$.

\begin{figure}[tbh]
\begin{center}
  \includegraphics[trim=0cm 1.5cm 0cm
    0cm,width=0.45\textwidth]{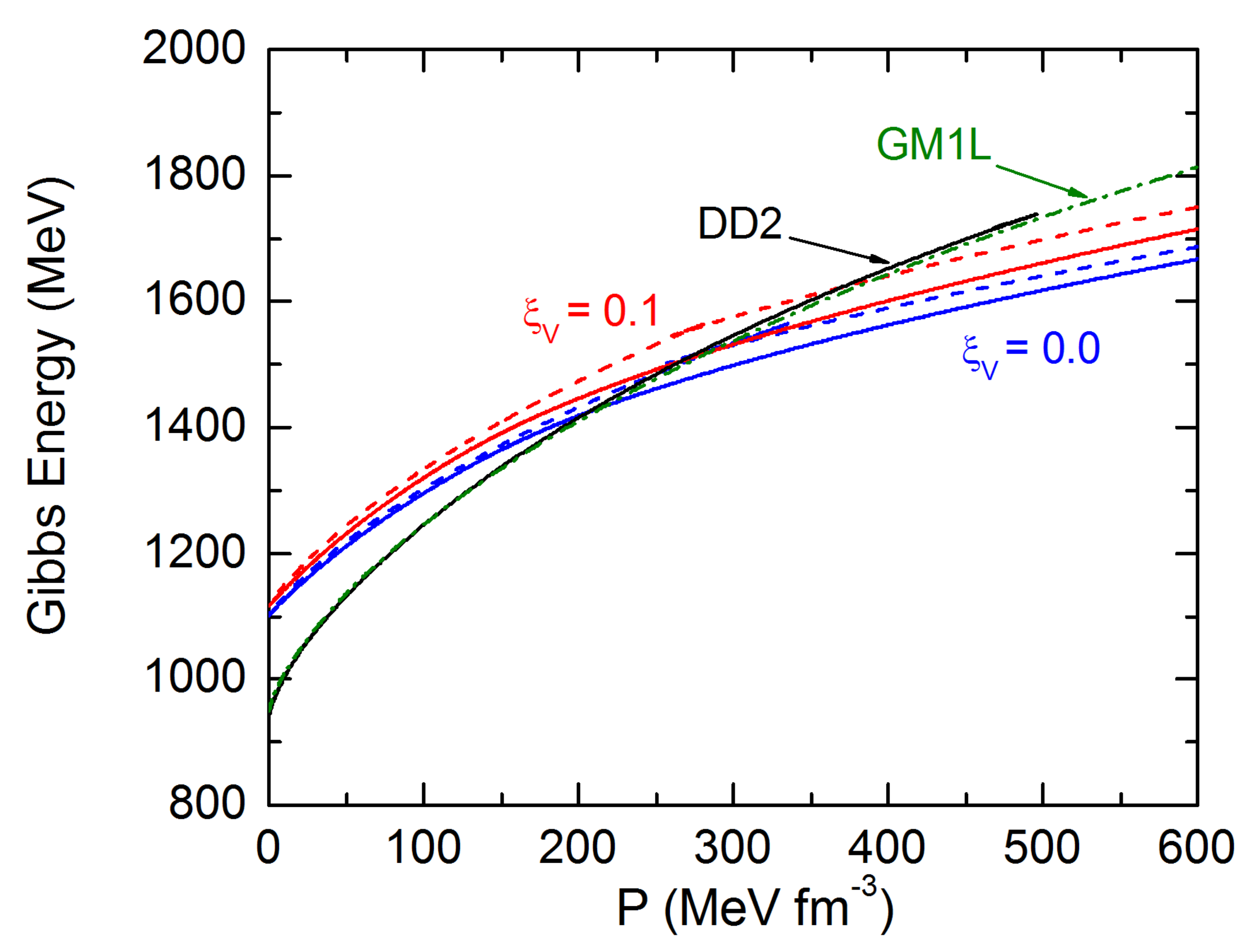}
  \includegraphics[trim=0cm 1.5cm 0cm
    0cm,width=0.45\textwidth]{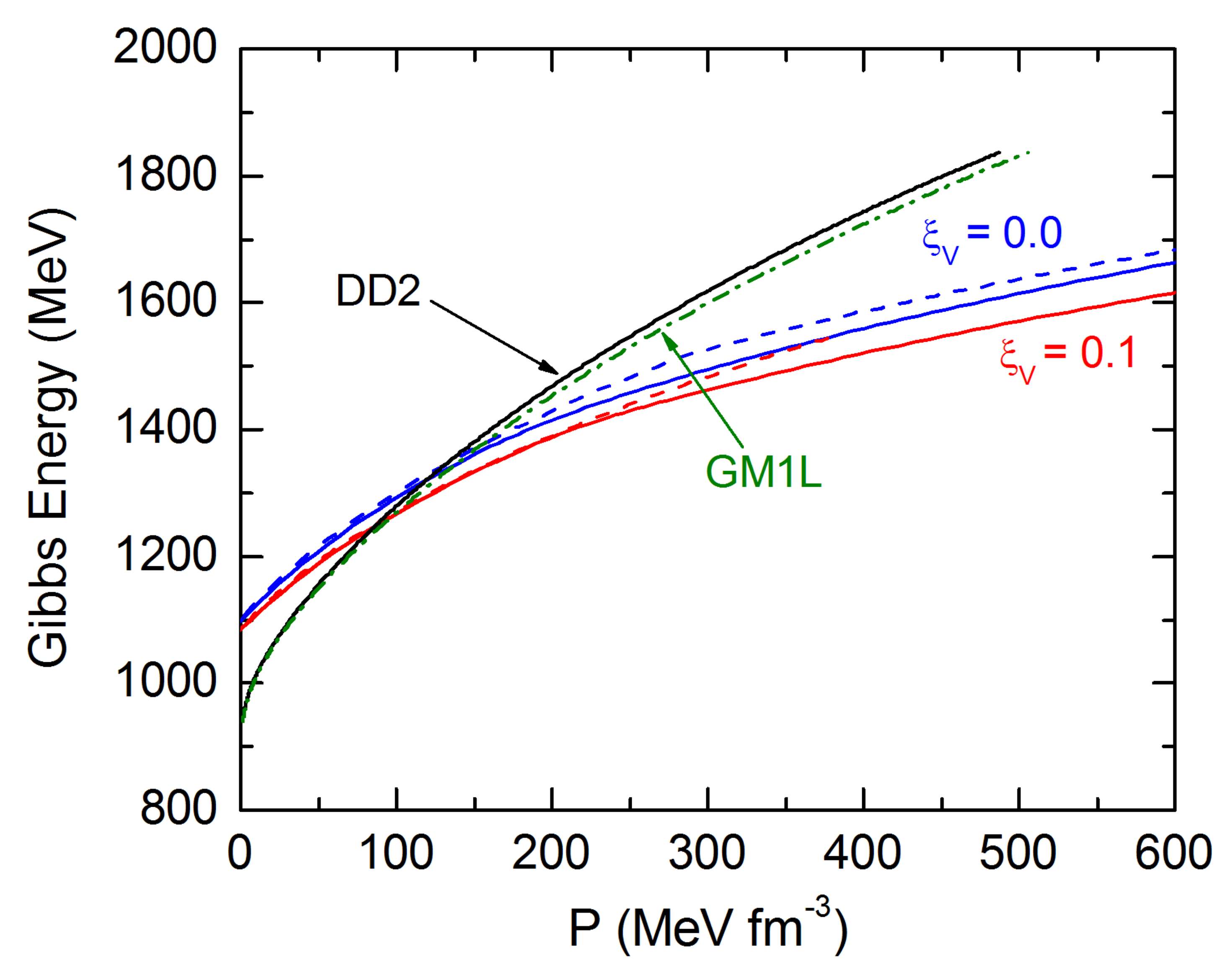}
\end{center}
\caption{Gibbs energy as a function of pressure.  The~ black and green
  curves refer to the Gibbs energy of hadronic matter computed for
  parameter sets DD2 and GM1L, respectively.  Quark matter is treated
  with the  {PNJL} model. The~'t Hooft quark flavor mixing term is
  included (absent) in the solid (dashed) quark matter curves.
  The~left panel is for matter at $T=0$ MeV, the~right panel for
  matter at $T=25$ MeV.}
  \label{fig:gibbs}
\end{figure}

\begin{figure}[tbh]
\centering
\includegraphics[width=0.6\textwidth]{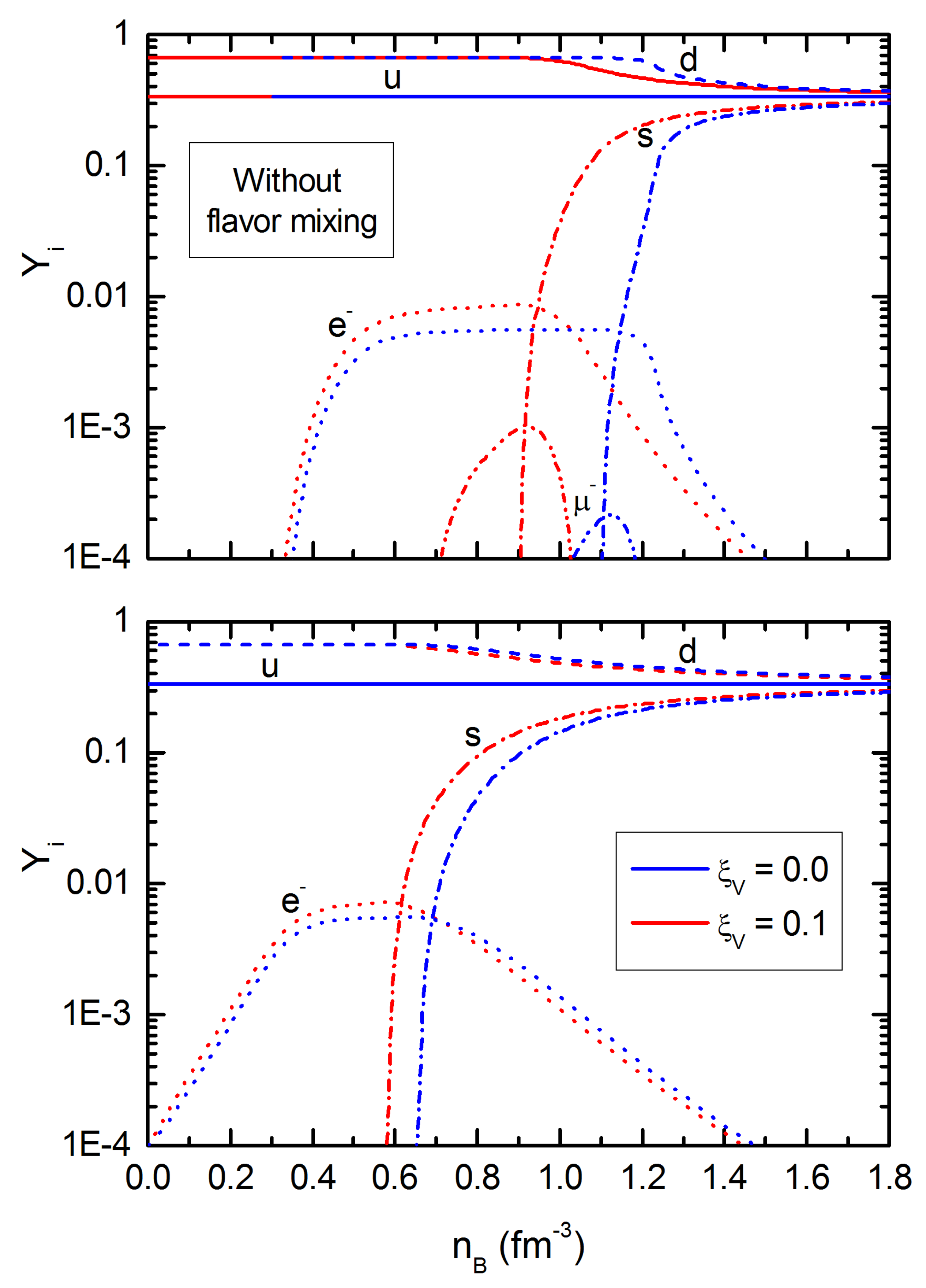}%\\ %\vspace{.5cm}
\caption{Impact of the vector interactions among quarks on the
  composition of cold quark matter. The~results shown in the top panel
  account for the 't Hooft term.}
\label{fig:thooft}
\end{figure}

To construct the hadron-quark phase transition we, as~described in
Section~\ref{sec:structure}, the~Gibbs condition $G_H(P, T, \{ \phi_H
\}) = G_Q (P, T, \{ \phi_Q \})$, where $G_{H}$ respectively $G_{Q}$
denote the Gibbs free energy per baryon of the hadronic ($H$) and the
quark ($Q$) phase at a given pressure and transition temperature. The~Gibbs energy of each phase ($i=H,Q$) is given by
\begin{eqnarray}
G_i (P,T) = \sum_j \frac{n_j}{n} \mu_j \, ,
\end{eqnarray}
where the sum over $j$ is over all the particles present in each
phase. Figure~\ref{fig:gibbs} shows the Gibbs energy of hadronic
matter and quark matter as a function of pressure. A~sharp
(Maxwell-like) first-order phase transition is assumed.  Values of
$\xi_{\mbox{v} } = 0$ and $0.1$ have been chosen for the ratio of the
vector-to-scalar quark coupling constant to demonstrate the dependence
of the Gibbs energy on the repulsion among quarks. The~points where
the curves cross each other mark the location of phase equilibrium
between hadronic and quark~matter.
Figure~\ref{fig:thooft} shows the composition of cold quark matter in chemical
equilibrium computed for  $\xi_{\mbox{v} } = 0$ and $0.1$.
It turns out that the 't Hooft term responsible for quark flavor mixing
alters the quark population significantly, which, in~turn, severely
changes the populations of leptons and~muons.

\begin{figure}[tbh] 
\centering
\includegraphics[trim=0cm 0.5cm 0cm 0cm,width=0.7\textwidth]{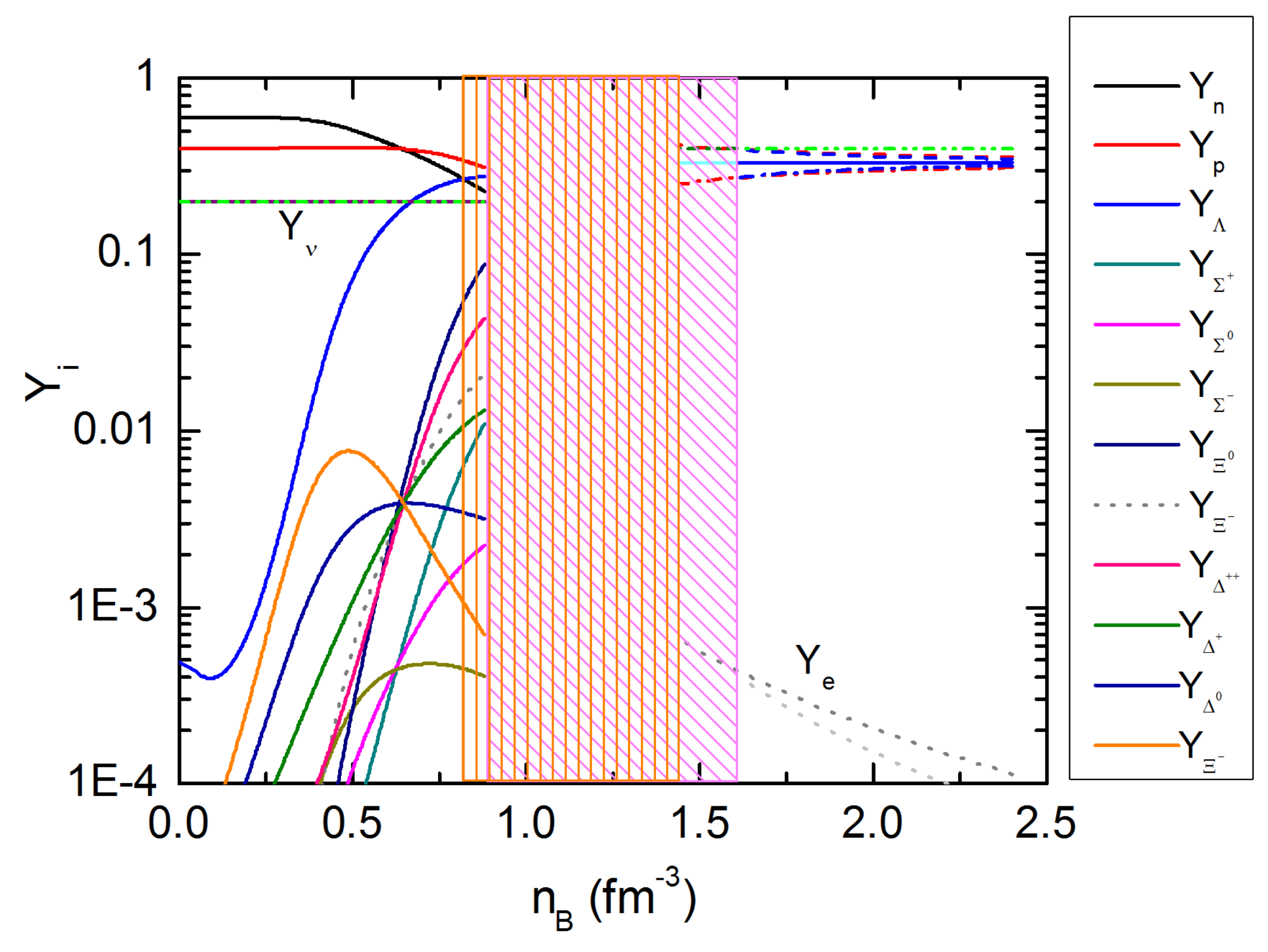}% \\ %\vspace{.5cm}
\caption{Particle population of proto-neutron star matter at a
  temperature of \mbox{$T=25$ MeV}, with~(vertical orange lines) and
  without (hatched magenta lines) quark flavor mixing.}
\label{fig:population25}
\end{figure}

In Figure~\ref{fig:population25}, we show the hadron quark composition
of hot neutron star matter.  Neutrinos are included. This composition,
therefore, corresponds to matter that may exist in the cores of
proto-neutron stars. The~shaded areas indicate a sharp (Maxwell)
hadron-quark phase transition, computed for the DD2 nuclear
parametrization. Three-flavor quark matter is described by the 
local  {PNJL}
model described~above.
As can be seen, quark flavor mixing shifts the hadron-quark phase
transition toward a smaller baron number density. This trend is to be
expected from the results for the particle densities shown in
Figure~\ref{fig:thooft}, which shows that flavor mixing shifts the
presence of quarks to lower~densities.

\section{Summary and Future~Prospects}

The purpose of this paper is to provide a short overview of the role
of quark matter for neutron star matter at different
temperatures. As~shown, the~hadron-quark composition of such matter
depends critically on temperature. For~proto-neutron star matter
neutrinos are to be included in the treatment too, which also impacts
the various particle fractions present at a given temperature and
density. If~hadron-quark matter exists in cold neutron star, it may
form a Coulomb lattice. As~demonstrated here (for a non-local version
of the NJL model), this lattice would be made almost entirely of
spherical quark blobs. The~size of the lattice depends on the central
density of a neutron star, which, most~interestingly, links the size
of the lattice to the spin-frequency of a neutron star. We find that
the lattice could be produced during spin-down of massive pulsars,
once their spin frequency has dropped below around 300 Hz ($\sim 25$\%
of the Kepler frequency).  The~study of quark matter with a local
 {PNJL} shows that the 't Hooft quark flavor mixing term leads to
non-negligible changes in the particle composition of (proto-) neutron
stars made of hadron-quark~matter.

We have analyzed the role of the 't Hooft mixing term, associated with
the axial U(1) symmetry anomaly of QCD, for~quark matter that is
locally electrically charge neutral, color neutral, and~in chemical
equilibrium (i.e., matter under neutron star matter conditions).  This
symmetry is expected to be restored (not necessarily together with
chiral symmetry) at high temperature, which implies that the mixing
coupling constant $H \rightarrow 0$.  For~high-density matter as
existing in the cores of neutron stars, the~mixing (six-point
interaction) term is less relevant than the four-point scalar and
vector interaction terms (see Equations~(\ref{L3}) and
(\ref{grandpotential})). At~the mean-field level, the~axial U(1)
symmetry anomaly of QCD does not change the thermodynamics of the
quarks~\cite{Blaschke:2005}.

Our results show that the inclusion of the mixing term lowers the
pressure at which the quark-hadro phase transition occurs.  It will
be interesting to explore how the mixing term modifies the possible
formation of a pasta phase in the hadron-quark mixed phase
if color super-conductivity is taken into account, since the mixing
term can destabilize a color superconductor by preventing the
formation of diquark condensates~\cite{Steiner:2005}.

The existence of a solid hadron-quark pasta phase in the core of a
neutron star may be linked to sudden pulsar spin-ups (glitches) and
the subsequent healing of the pulsar period. A~very prominent example
of a pulsar that displays regular glitch activity is PSR
J0537--6910. The~glitch rate of this pulsar is 3.2 yr$^{-1}$
\cite{Ferdman:2018}.  A~link between pulsar glitches and solid
neutron-matter cores inside of neutron stars has been pointed out may
years ago already ~\cite{Pines:1972a,Pines:1972b,Canuto:1973}. This
suggests that one should investigate the connection between a solid
hadron-quark pasta phase and the pulsar glitch
phenomenon. In~particular, it needs to be studied whether or not the
sudden release of elastic energy stored in a solid hadron-quark pasta
phase could explain macro-glitches (which are related to sudden
changes in the rotational frequency of a pulsar on the order of
$\Delta\Omega/\Omega \sim 10^{-6}$) at the rate (months to years) at
which pulsar glitches are typically observed. Another issue that needs
to be investigated concerns the role of neutron superfluidity.  Since
neutron stars with and without hadron-quark matter cores contain
nuclear crusts with identical compositions (but varying thickness),
both types of stars possess nuclear crusts containing $^1{\rm S}_0$
neutron superfluids. The~situation is less clear for the neutrons in
the cores of both types of stars. It needs to be studied if both types
of stars may possess $^3{\rm S}_1$ neutrons superfluids in their
cores, or~whether this phase is restricted to neutron stars without
quark matter cores.

Gravitational-wave detectors such as LIGO and VIRGO have opened up a
new observational window on the inner workings of neutron stars. This
is particularly the case for the neutron stars in the elliptic galaxy
NGC 4993 that produced the gravitational-wave event GW170817 observed
with LIGO and VIRGO in 2017.  The~detection of the gravitational and
electromagnetic waves produced by the coalescence of these neutron
stars has offered an unprecedented opportunity for imposing
constraints on the quark-hadro composition of neutron stars and the
equation of state associated with such objects (see
Refs.~\cite{Hanauske:2017,Hanauske:2018JAA,Hanauske:2019},
and~references therein).

\vspace{6pt}

\authorcontributions{The authors contributed equally to the
  theoretical and numerical aspects of the work presented in this~paper.}

\funding{M.G.O.\ and G.A.C.\ thank UNLP and CONICET for
  financial support under Grants G157, X824 and PIP-0714. F.W.\ is
  supported by the National Science Foundation (USA) under Grant
  PHY-1714068.}

\conflictsofinterest{The authors declare no conflict of~interest.}

\reftitle{References}

% macros used by ADS Database BiBTeX entries:
% see http://adsabs.harvard.edu/abs_doc/aas_macros.sty
\newcommand{\apj}{Astrophys. J.\ }
\newcommand{\prd}{Phys. Rev. D \ }
\newcommand{\prc}{Phys. Rev. C \ }
\newcommand{\apjl}{Astrophys. J. Lett.\ }
\newcommand{\mnras}{Mon. Not. R. Astron. Soc.\ }
\newcommand{\aap}{Astron. Astrophys.\ }
\newcommand{\jcap}{Journal of Cosmology and Astroparticle Physics\ }

%\externalbibliography{yes}

%\bibliography{universe-2019}
%\reftitle{references}

\end{document}